\documentclass[lettersize,journal]{IEEEtran}

{ \vspace{-0.15cm}%
    \small\noindent{\bfseries Availability of Data and Material:}\par%
    \noindent\ignorespaces}%
{ \par\noindent%
\ignorespacesafterend }%

\AtBeginDocument{%
  }

\usepackage{amsmath,amsfonts}
\usepackage{algorithmic}
\usepackage{algorithm}
\usepackage{array}
\usepackage[caption=false,font=normalsize,labelfont=sf,textfont=sf]{subfig}
\usepackage{textcomp}

\usepackage{stfloats}
\usepackage{url}
\usepackage{verbatim}

\usepackage{graphicx}
\usepackage{comment}

\usepackage{makecell}
\usepackage{rotating}
\usepackage{booktabs}
\usepackage{siunitx}
\usepackage{colortbl}
\usepackage{dsfont}
\newcommand{\argmin}[1]{\underset{#1}{\text{argmin}}\,}
\usepackage[switch]{lineno}
\usepackage{placeins}
\begin{document}

\title{Quantification of Residential Flexibility Potential using Global
Forecasting Models}

\author{Lorenzo Nespoli, Vasco Medici
        % <-this % stops a space
%\thanks{Manuscript received MM DD, XXXX; revised MM DD, XXXX.}

\thanks{This work was financially supported by the Swiss Federal Office of Energy (ODIS – Optimal DSO dISpatchability, SI/502074) and the Swiss National Science Foundation under NCCR Automation (grant agreement 51NF40 180545), and supported by IEA Annex 82 "Energy Flexible Buildings Towards Resilient Low Carbon Energy Systems".} 
\thanks{Lorenzo Nespoli and Vasco Medici are with ISAAC, DACD, SUPSI, Mendrisio, CH (email lorenzo.nespoli@supsi.ch, vasco.medici@supsi.ch). Lorenzo Nespoli is with Hive Power SA, Manno, CH}% <-this % stops a space
}
\maketitle

\begin{abstract}
This paper proposes a general and practical approach to estimate the economic benefits of optimally controlling deferrable loads in a Distribution System Operator's (DSO) grid, without relying on historical observations. We achieve this by learning the simulated response of flexible loads to random control signals, using a non-parametric global forecasting model. An optimal control policy is found by including the latter in an optimization problem. We apply this method to electric water heaters and heat pumps operated through ripple control and show how flexibility, including rebound effects, can be characterized and controlled. Finally, we show that the forecaster's accuracy is sufficient to completely bypass the simulations and directly use the forecaster to estimate the economic benefit of flexibility control.
\end{abstract}


\section{Introduction}\label{sec:motivations}
Flexibility is a term used to describe the ability of electric loads or distributed energy resources (DERs) to shift their consumption or production in time. Flexibility in distribution or transmission grids can increase grid resilience, reduce maintenance costs, lower distribution losses, and smooth and increase the predictability of the demand profile \cite{j_cochran_et_al_flexibility_nodate,babatunde_power_2020,mohandes_review_2019}. Flexibility services usually require aggregating flexible residential customers into pools that reach a given "critical mass" \cite{eid_aggregation_2015,parvania_optimal_2013}. In most cases, aggregation requires controlling heterogeneous types of devices \cite{ghaemi_optimal_2019} (e.g., heat pumps, electric boilers, EVs, PVs), running different types of onboard controllers, (e.g., rule or heuristic-based, model predictive control, etc..). This condition restricts the kind of viable control methods for pooling flexibility. Some protocols, such as OSCP \cite{noauthor_oscp_nodate}, envisage intermediate actors optimizing flexibility pools by means of a global control signal, delegating the complexity of low-level control to a flexibility provider \cite{portela_oscp-open_2015, biegel_aggregation_2014}. Currently, the most used control method is ripple control \cite{chen_ripple-based_2016}, using frequency-sensitive relays to shut down flexible devices. Aggregating loads in control pools reduces uncertainty in the total amount of actuated flexibility \cite{ponocko_forecasting_2018}; yet, communicating instant flexibility may prove insufficient for optimal dispatch. Frequently, deactivating a cluster of energy-intensive devices might trigger a "rebound effect" in the overall load once they are reactivated \cite{cui_evaluation_2018}. This effect can create an unintended spike in peak demand, a factor that should be taken into account when optimizing the overall power profile.
 
%% On the other hand, to train the response of a dynamic tariff we must revert to a pure data-driven approach, since in this case we assume that the influence of the tariff on the aggregated power is due to unknown controllers or human decisions, whose behaviour we cannot accurately simulate. 

\IEEEpubidadjcol
\subsection{Related work}
Flexibility research has gained prominence in recent publications. For example, the International Energy Agency's (IEA) Annex 67 \cite{jensen_iea_2017} focuses on using building flexibility for grid control, and Annex 82 \cite{noauthor_iea_nodate} examines its quantification for utilities and DSOs. Some publications are mostly focused on the characterization of flexible devices \cite{six2011exploring, chen_measures_2018, balint_determinants_2019, junker_characterizing_2018, nuytten_flexibility_2013} while others mostly explore its exploitation in the context of demand side management and demand response, under the hypothesis of a known, observable and directly controllable system \cite{petersen_taxonomy_2013,de_coninck_quantification_2016,oldewurtel_towards_2013}.  For example, in \cite{oldewurtel_towards_2013},  \cite{de_coninck_quantification_2016} and \cite{reynders_generic_2017}, this is achieved for thermally activated building system (TABS) and heat pumps (HPs).
Our work is mostly related to simulation-based flexibility assessment of partially observable and indirectly controllable systems. This setting resembles the current operational conditions of electrical grids: DSOs usually can only rely on smart meters' relays for load actuation, and temperature readings are not available. Similar conditions were considered in \cite{fischer_model-based_2017}, where authors assessed the energy flexibility potential of a pool of residential smart-grid-ready HPs (i.e., with an internal controller reacting to a discrete signal indicating if they have to consume more, less, or shut down) by means of bottom-up simulations. Similarly, in \cite{muller_large-scale_2019}, authors predicted the energy consumption of a group of 300 HPs controlled via binary throttling signals. In \cite{valles_probabilistic_2018}, the authors trained a forecaster on periods in which demand-response is not active to quantify the flexibility associated with a pool of customers under a price-and-volume schema. This approach was possible due to the sparsity of actuation events, allowing to separate baseline and activation periods. Our work is also related to inverse optimization of price signals, which was first introduced in \cite{corradi_controlling_2013}. The idea is that assuming that some buildings use a price-dependent (but unknown) controller, the DSO or an aggregator can try to reverse engineer the controllers by estimating approximate and invertible control laws by probing the system with a changing price signal; since the learned control laws are invertible, they can then be used to craft the optimal cost signal to provide a desired aggregate power profile. To show this, authors in \cite{corradi_controlling_2013} fitted an invertible online FIR model to forecast the consumption of a group of buildings as a function of a price signal and derive an analytic solution for an associated closed-loop controller. The concept was then demonstrated by means of simulations on 20 heat-pump-equipped households. The authors of \cite{junker_characterizing_2018} used the same concept to fit a linear model linking prices and the load of a cluster of price-sensitive buildings. The authors then proposed to characterize flexibility extracting parameters from the model response. They also proposed to estimate the expected savings of a given building by simulating its model twice, with and without a price-reacting control. A similar approach was proposed in \cite{junker_stochastic_2020}, where authors identified a general stochastic nonlinear model for predicting energy flexibility coming from a water tower operated by an unknown control strategy. The fitted model is then used in an optimization loop to design price signals for the optimal exploitation of flexibility. Authors in \cite{yin_long-term_2022} used the same method to find price signals to best meet flexibility requests using an iterative method.

\subsection{Contributions}
Opposed to the approaches presented in the reviewed literature, which employ simple invertible models to estimate flexibility \cite{junker_characterizing_2018, junker_stochastic_2020, corradi_controlling_2013}, we propose to train global forecasters or metamodels, based on boosted trees on simulated data to predict both the controlled and uncontrolled power of flexible devices. This allows conditioning the response on categorical variables, such as the number of controlled devices of different types and past binary control signals generated by ripple control or throttling. This latter ability allows the use of the forecaster as a surrogate model of the simulation inside a control loop. We also show that global models provide sufficient accuracy to bypass the simulations and to perform the same kind of what-if analysis presented in \cite{fischer_model-based_2017}. This is possible because we are only interested in the aggregated power of the controlled devices, which has a much lower dimensionality than all the simulated states and signals. The method we propose can be used to assess the power response of groups of flexible devices from day zero by means of simulations but can also be applied to real controlled systems (for which it is not possible to retrieve a baseline response) by augmenting the training set using observations from the field. In section \ref{sec:modeling}, we show that the modeling and simulation phase needed to create a training set for the metamodel only requires statistical information, which is usually publicly available. In section \ref{sec:oracle}, we present a method to predict energy flexibility using a global forecasting model. We conduct an ablation study in which we suggest various training methodologies. These findings indicate that incorporating concepts of energy imbalances throughout the prediction horizon and crafting a training set from scenarios exhibiting orthogonal penetrations based on device types enhances the accuracy of forecasts. In \ref{subsec:rebound}, we use the metamodel to characterize flexibility and rebound effects, allowing us to answer complex questions like: How does the controlled device mix influence flexibility? And, how many kWh, at which power level, could be deferred?  In section \ref{sec:optimization}, we describe how the metamodel can be used to optimize the available flexibility. In section \ref{subsec:dynamic_grouping}, we propose a dynamic grouping strategy to ensure that the thermal comfort constraints of end users with an HP are never violated. Finally, in section \ref{sec:emulation}, we study the accuracy of the metamodel when used to optimize flexible devices. For the analyzed use case, we show that the metamodel is accurate enough to completely bypass the simulation, allowing us to use it for both simulation and control. 

\section{Problem statement and system description}\label{sec:modeling}
Our objective is to evaluate the flexibility potential of residential customer groups in response to a force-off control signal $s$.
Our approach involves learning a computationally-effective meta-model based on a detailed, white-box simulation of flexible devices, and incorporating this model within an optimal control loop to minimize operational costs.\\
We consider the setting in which a DSO plans a control signal $s \in \mathds{R}^{96}$ with a 15-minute resolution for the next day. In our simulations, the signal planning occurs every day at midnight, covering the subsequent 24 hours.
We restrict this study to two flexible devices, HPs and electric water heaters (EHs). We simulated the following heating system configurations:
\begin{enumerate}
    \item HP: in this configuration, both space heating and domestic hot water (DHW) are provided by an HP.
	\item EH: in this case, the EH is just used to provide DHW, while the space heating is not modeled, the latter being considered to be fueled by gas or oil.
\end{enumerate} 
A detailed mathematical description of the building thermal model, stratified water tanks, HP, and heating system model is provided in appendix \ref{annex_1}. To validate our methodology, we conducted simulations reflecting typical device usage and overall power consumption for a DSO in the Swiss canton of Ticino. Appendix \ref{annex_2} lists the data sources used to configure the simulated devices. Within this region, our analysis included 2670 buildings with installed HPs and 1750 with EHs, possessing a total nominal electrical capacity of 12.5 MW and 7.7 MW, respectively.

\section{Global forecasting modes for flexibility simulation and control}\label{sec:oracle}

We start considering a single group of simulated flexible devices. We define a dataset $\mathcal{D}_s = \{(x_t, y_t^f)_{t=1}^N\}$ of input-output tuples, where $x_t \in \mathds{R}^{n_f}$ is a set of $n_f$ features, including past and future values of the control signal $s$ sent to the group of devices, while $y_t^f \in \mathds{R}^H$ being their aggregated power profile for the next $H$ steps ahead. We want to use $\mathcal{D}_s$ to train a forecaster, or meta-model, $f(x, \theta):  x_t \rightarrow \hat{y}_t^f$.

\subsection{Dataset generation}\label{subsec:dataset_generation}
\begin{figure}[!h]
	\centering
	\includegraphics[width=1\columnwidth]{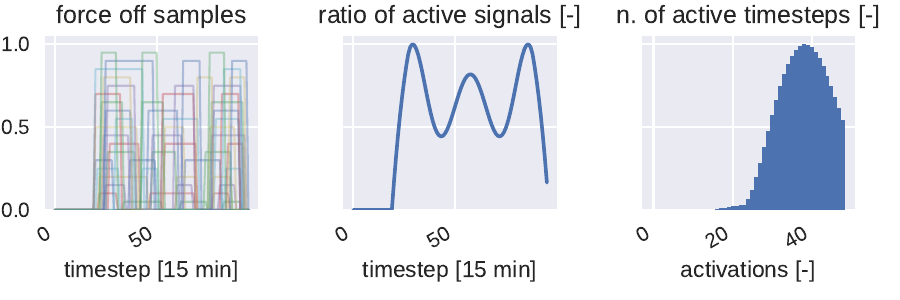}
	\caption{Left: a random sample of daily scenarios for the force-off signal. Center: ratio of active signals for a given timestep of the day. Right: distribution of the number of active timesteps among all possible scenarios.}
	\label{fig:scenarios}
\end{figure}

The dataset is built from a one-year simulation in which devices were controlled using a random control policy and a one-year uncontrolled simulation; this is opposed to simulating tuples of controlled and uncontrolled cases starting from the same system's state. The latter approach is more complicated, requiring resetting the simulation states each time; furthermore, it cannot be used when gathering data from real systems. To build the control signal $s$ for the controlled year, we generated all possible daily random signals respecting specific criteria, such as a daily mandated minimum period for sustained state and a capped number of daily activations; these criteria are reported in table \ref{tab:scenarios}. Using a 15-minute time-step will require generating ex-ante $2^{96}$ signals. For this reason, we used a dynamic programming approach, filtering out incompatible scenarios on the run, as they are sequentially generated. Figure \ref{fig:scenarios} shows a sample of the resulting force-off signals, the ratio of scenarios in which the force-off signal $s$ is active as a function of time-step, and the distribution of the total steps in which the force-off signal is on. 

\begin{table}[h]
\caption{Parameters used to generate all possible daily force-off signals\label{tab:scenarios}}
	\center
	\begin{tabular}{cc}
		\toprule
		parameter            & value \\
		\hline
		force off max steps       & 96                                                              \\
		\rowcolor[gray]{.95}  min constant period & 8  (2H) \\
		max number of switches         & 6                                                     \\
		\rowcolor[gray]{.95}  max on steps            & 48 (12H) \\ 
        nightly uncontrolled period & 20 (5H)\\
        \bottomrule
	\end{tabular}
\end{table}

Instead of training several metamodels using datasets with different numbers of HPs and EHs, we follow a common approach from forecasting literature and train a single global model by crafting datasets of different penetration scenarios and using them to create a single dataset. We build the final dataset following these steps: 
\begin{enumerate}
	\item we build penetration scenarios by grouping a subset of the simulated buildings, from which the aggregated power $y_t^f$ is retrieved. A dataset is then built for each penetration scenario, picking at random $k\%$ observations from the simulated years. We sampled a total of 100 penetration scenarios and used $k=20$, for a total length of the dataset of 40 equivalent years.
	\item we retrieve metadata describing the pool of buildings for each penetration scenario. Metadata includes the total number of each kind of device, the mean thermal equivalent transmittance (U) of the sampled buildings, and other parameters reported in table \ref{tab:metadata}. We further augment the dataset with time features such as the hour, the day of the week, and the minute of the day of the prediction time.  
    \item Augment each penetration scenario dataset through transformations and lags of the original features, as reported in table \ref{tab:training_set}, to obtain $\mathcal{D}_{s}$. 
    \item Retrieve the final dataset by stacking the penetration scenario datasets $\mathcal{D} = [\mathcal{D}_s]_{1:n_s}$  
\end{enumerate}

\begin{table}[]
\caption{Metadata used as features in the training set. Penetration scenario features describe the characteristics of the pool of simulated buildings and devices, while temporal features refer to the time of the prediction. Here $q_{z\%}$ stands for the $z\%$ quantile.\label{tab:metadata}}
	\center
	\begin{tabular}{ccc}
		\toprule
		penetration scenario features & temporal features \\
		\hline  
		\makecell{Sum, $q_{10\%}$ and $q_{90\%}$ \\ of the nominal powers of devices,\\ number of HPs and EHs and their ratio,\\
			Mean, $q_{10\%}$ and $q_{90\%}$ of thermal resistances,\\Mean, $q_{10\%}$ and $q_{90\%}$ of thermal capacities} & \makecell{hour, day of week, \\ minuteofday}\\
        \bottomrule
	\end{tabular}
\end{table}

\begin{table}[]
\caption{Continuous variables, transformations and lags passed as features to the metamodel. Meteorological information consists of temperature and global horizontal irradiance measurements. \label{tab:training_set}}
    \begin{tabular}{ccc}
		\toprule
		signals & transformation & lags\\
		\hline
		$s$       & \makecell{shifts(15m) \\ mean(3h), mean(6h)} & \makecell{-95,...96 \\  1...96 } \\
		\rowcolor[gray]{.95}$y_t^f$, meteo&\Gape[0pt][2pt]{\makecell{shifts(15m) \\ mean(1h)}}&\makecell{-4,..0 \\ -168..-144, -24...0}\\
		meteo & mean(1h) & 1..24 \\
        \bottomrule
    \end{tabular}
\end{table}

\subsection{Model description}\label{subsec:model}

The metamodel is a collection of multiple-input single-output (MISO) LightGBM regressors \cite{noauthor_welcome_nodate} predicting $y_t^f$ at a different step-ahead. The alternative to a collection of MISO models is training just one MISO model after augmentation of the dataset with a categorical variable indicating the step ahead being predicted. This option was discarded due to both memory and computational time restrictions. For our dataset, this strategy requires more than 30 GB of RAM. Furthermore, training a single tree for the whole dataset requires more computational time than training a set of MISO predictors in parallel (on a dataset that is 96 times smaller). We recall that the final dataset is composed of 100 scenarios differing in the set of buildings composing the aggregated response to be predicted. This means that removing observations at random when performing a train-test split would allow the metamodel to see the same meteorological conditions present in the training set. To overcome this, the training set was formed by removing the last 20\% of the yearly observations from each penetration scenario dataset $\mathcal{D}_s$. That is, the training-test split is done such that the training set contains only observations relative to the first 292 days of the yearly simulation.

A hyper-parameter optimization is then run on a 3-fold cross-validation over the training set; this means that each fold of the hyper-parameter optimization contains roughly 53\% of $\mathcal{D}$. The tuned hyper-parameters are just the learning rate and the number of estimators for the LightGBM regressors; the parameters are kept fixed for all 96 models predicting the various step-ahead. We used a fixed-budget strategy with 40 samples, using the \texttt{optuna} python package \cite{noauthor_optuna_nodate} implementation of the tree-structured Parzen estimator \cite{ozaki_multiobjective_2020} as a sequential sampler.

\subsection{Ablation studies}\label{subsec:ablation}
We performed an ablation study to see the effectiveness of different sampling strategies (point (1) of the dataset-building methodology described in the previous section) and model variations. 
\paragraph{Sampling schemes}
To generate the final dataset, we tested two different sampling schemes for producing the penetration scenarios. In the first strategy, the total number of controllable devices is increased linearly, picking randomly between households with an HP or an EH. In the second strategy, the number of the two controllable classes of devices is increased independently, co-varying the number of HPs and EHs in a cartesian fashion. 
\begin{figure}[!h]
	\centering
	\includegraphics[width=1\columnwidth]{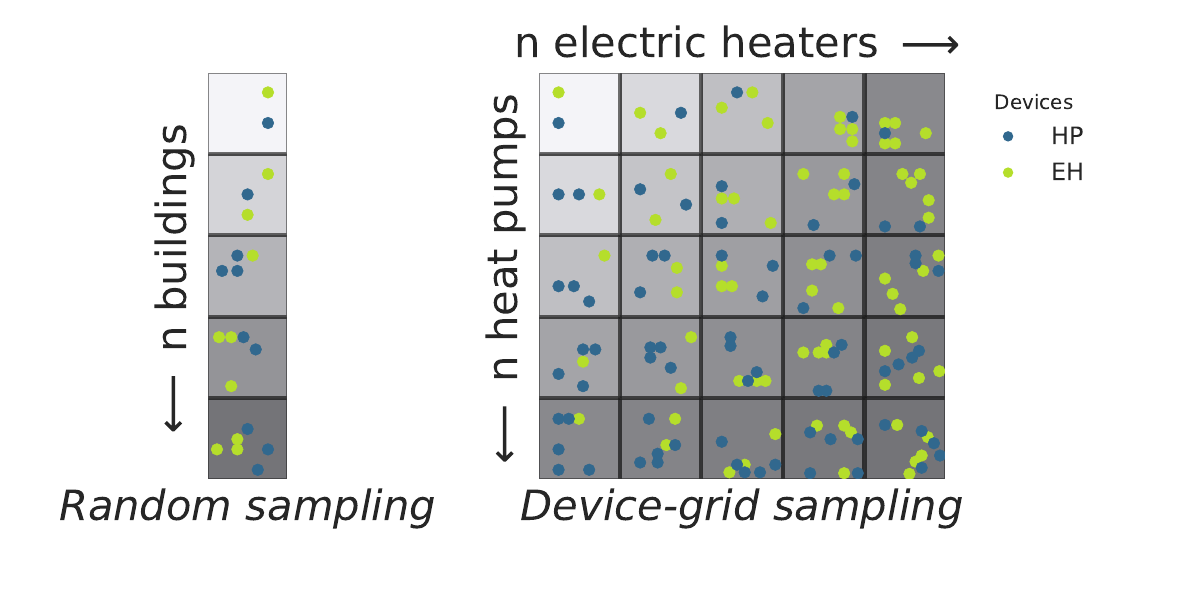}
	\caption{Sampling strategies for building the final training set. Left: the total number of controllable devices is increased linearly, picking randomly between households with an HP or an EH. Left: the number of controllable devices is increased by independently co-varying the number of HPs and EHs.}
	\label{fig:sampling_scheme}
\end{figure}

\paragraph{Energy unbalance awareness}

To enhance the accuracy of the metamodel, a physics-informed approach involving energy imbalance is proposed. This method utilizes the metamodel to simulate the system's response under two conditions: with the actual control signal $s$ and with a zeroed control signal. By subtracting these responses, we quantify the system's 'energy debt' at each timestep. This physics-based insight is crucial for improving predictions of future states. To test this hypothesis, we developed a secondary model where a set of regressors first forecasts the system response for future steps under both scenarios. The resultant energy imbalances from these predictions serve to enrich the training dataset. Subsequently, another set of regressors is trained on this augmented dataset, employing this physics-informed strategy during both training and prediction phases.

\begin{figure}[!h]
	\centering
	\includegraphics[width=1\columnwidth]{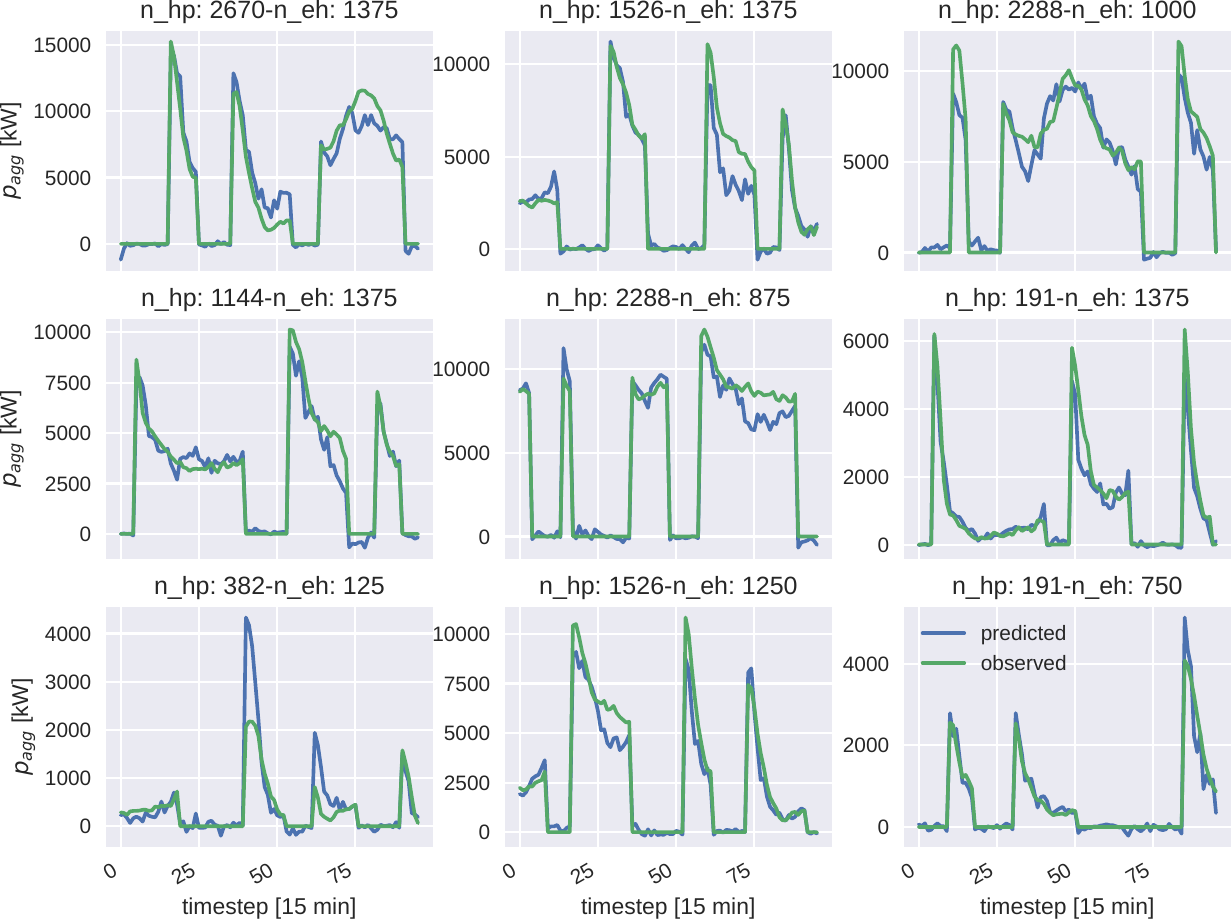}
	\caption{Random example of day-ahead metamodel's forecasts, for different numbers of HPs and EHs, where the force off was activated at least once, for the energy-aware metamodel trained using the grid sampling strategy}
	\label{fig:ctrl_examples}
\end{figure}

In total, we compared four distinctive configurations, comprising the two models and the two sampling strategies. Figure \ref{fig:ctrl_examples} provides representative examples of predictions of the energy-aware metamodel trained using the grid sampling strategy, featuring varying counts of controlled heat pumps (HPs) and electric heaters. 

\begin{figure}[!h]
	\centering
	\includegraphics[width=1\columnwidth]{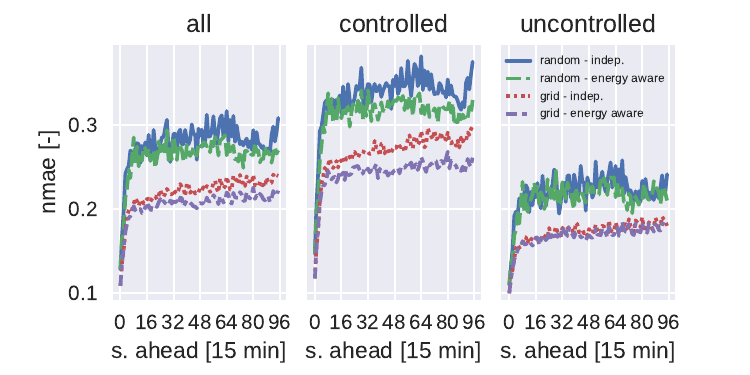}
	\caption{Performances for the four tested metamodels, in terms of nMAE as a function of the step ahead.}
	\label{fig:score_lineplot}
\end{figure}

\begin{comment}

Figure \ref{fig:heatmap_time} shows a heatmap illustrating the Normalized Mean Absolute Error (NMAE) as a function of the prediction time and step ahead. The four models under investigation presented the same qualitative patterns. They obtain low NMAE values for the first steps ahead and demonstrate superior predictive capability during nighttime hours, whereas the NMAE surges during peak periods. 

\begin{figure}[!h]
	\centering
	\includegraphics[width=0.45\textwidth]{figs/cartridge_results_power_grid_2023-01-27_18.pk_score_multisummary_hour.pdf}
	\caption{NMAE for the two tested models for the power oracle, when using grid samples, as a function of step-ahead and time of prediction.}
	\label{fig:heatmap_time}
\end{figure}
\end{comment}
Models performances can be better compared when plotting the average (over samples and prediction times) normalized Mean Absolute Error (nMAE) as a function of step ahead, as done in figure \ref{fig:score_lineplot}. The nMAE for the predictions generated at time $t$ is defined as:
\begin{equation}\label{eq:nmae}
    nMAE_t =  \frac{ \sum_{k=1}^{96} \vert y_{t+k}- \hat{y}_{t+k}\vert}{\sum_{k=1}^{96} \vert y_{t+k}\vert}
\end{equation}

The grid sampling scheme did indeed help in increasing the accuracy of the predictions w.r.t. the random sampling scheme for both the LightGBM models. Including the information about energy imbalances at each step ahead shows some benefits for both sampling strategies, at the expense of a more complex model. The accuracy improvement impacts only controlled scenarios, as demonstrated by comparing the second and third panels in figure \ref{fig:score_lineplot}. These panels show the scores obtained for instances where the force-off signal was activated at least once or never activated. This result aligns with our expectations. As an additional analysis, we studied the energy imbalance over the prediction horizon. For this analysis, we considered just the controlled cases in the test set. We define two relative energy imbalanced measures:
\begin{equation}
\Delta_{rel} E_t = \frac{\sum_{k=1}^{96} \hat{y}_{t+k}(s)-\sum_{k=1}^{96}y_{t+k}}{\sum_{k=1}^{96}y_{t+k}}
\end{equation}
\begin{equation}
\Delta_{rel}^{noctrl} E_t = \frac{\sum_{k=1}^{96} \hat{y}_{t+k}(s)-\sum_{k=1}^{96}\hat{y}_{t+k}(s_0)}{\sum_{k=1}^{96}y_{t+k}}
\end{equation}
where $y_t$ is the simulated power, $\hat{y}(s)$ is the power predicted by the metamodel with the control used in the simulation, and $\hat{y}(s_0)$ is the power predicted by the metamodel using a zero force off. We can interpret $\Delta_{rel} E_t$ as the relative error in the total energy needs w.r.t. the simulation and $\Delta_{rel}^{noctrl} E_t$  as the change in the energy consumption estimated by the metamodel if the pool of flexible devices were not controlled.  We removed from the comparison all the instances in which the force-off signal was activated in the last 5 hours of the day. In this case, part of the consumption will be deferred outside the prediction horizon, making the comparison meaningless.

\begin{figure}[!h]
	\centering
	\includegraphics[width=1\columnwidth]{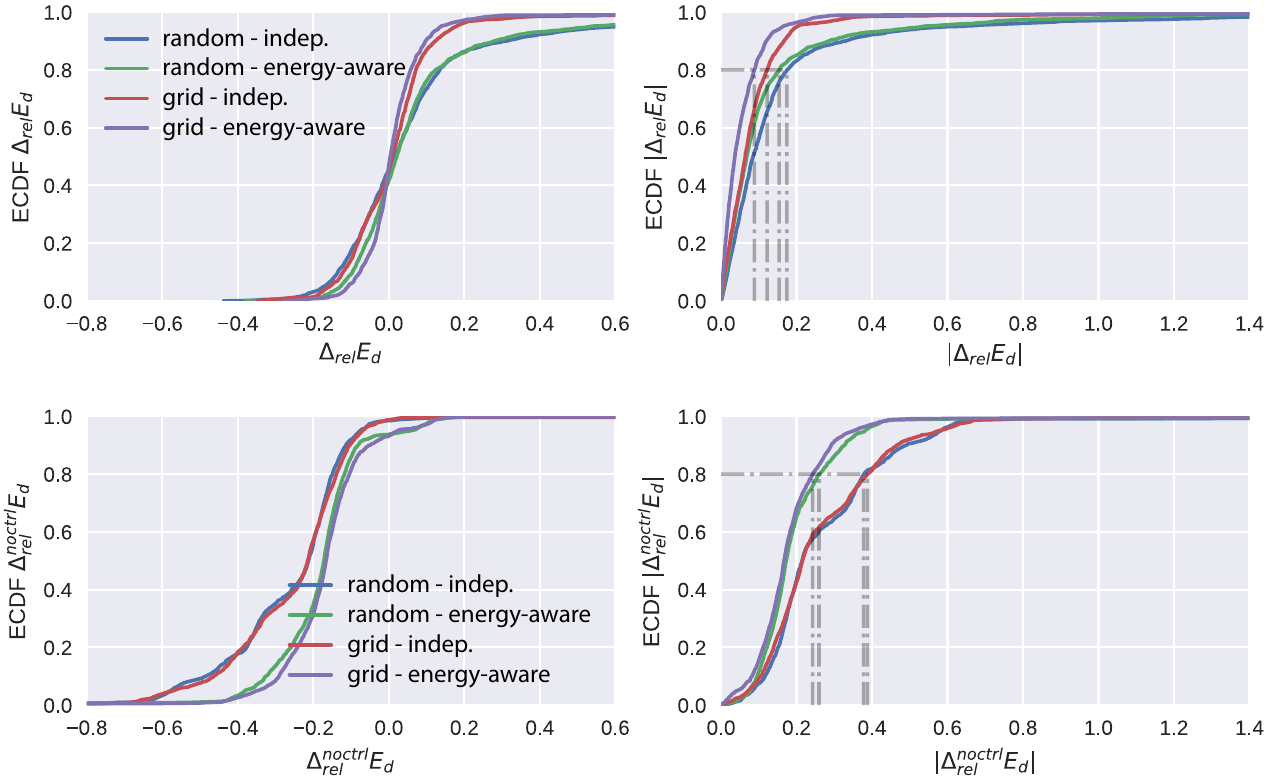}
	\caption{Left: cumulative distributions of the relative energy imbalance for different models. Right: empirical cumulative density functions of absolute relative energy imbalance for different models.}
	\label{fig:energy_unbalance}
\end{figure}

Looking at the first row of figure \ref{fig:energy_unbalance}, we see how the empirical cumulative distribution functions (ECDFs) of $\Delta_{rel} E_d$ and its absolute value (left and right panels) are closer to zero when the grid sampling strategy is applied. Also, using the energy-aware model helps in having a more precise prediction in terms of used energy over the prediction horizon. For all 4 models, 80 \% of the time, the relative deviation in the horizon energy prediction lies below 20\%. The second row of figure \ref{fig:energy_unbalance} reports the change in the forecasted energy consumption within the prediction horizon with and without control. It is reasonable to think that the consumption should approximately match since the force off usually just defers the consumption. In this case, the energy-aware models present a lower difference in the consumed energy. 

%\begin{figure}[!h]
%	\centering
%	\includegraphics[width=0.6\textwidth]{figs/energy_unbalance_hist.pdf}
%	\caption{}
%	\label{fig:energy_unbalance_hist}
%\end{figure}

\subsection{Characterization of the rebound effect}\label{subsec:rebound}
\begin{figure}[!h]
	\centering
	\includegraphics[width=1\columnwidth]{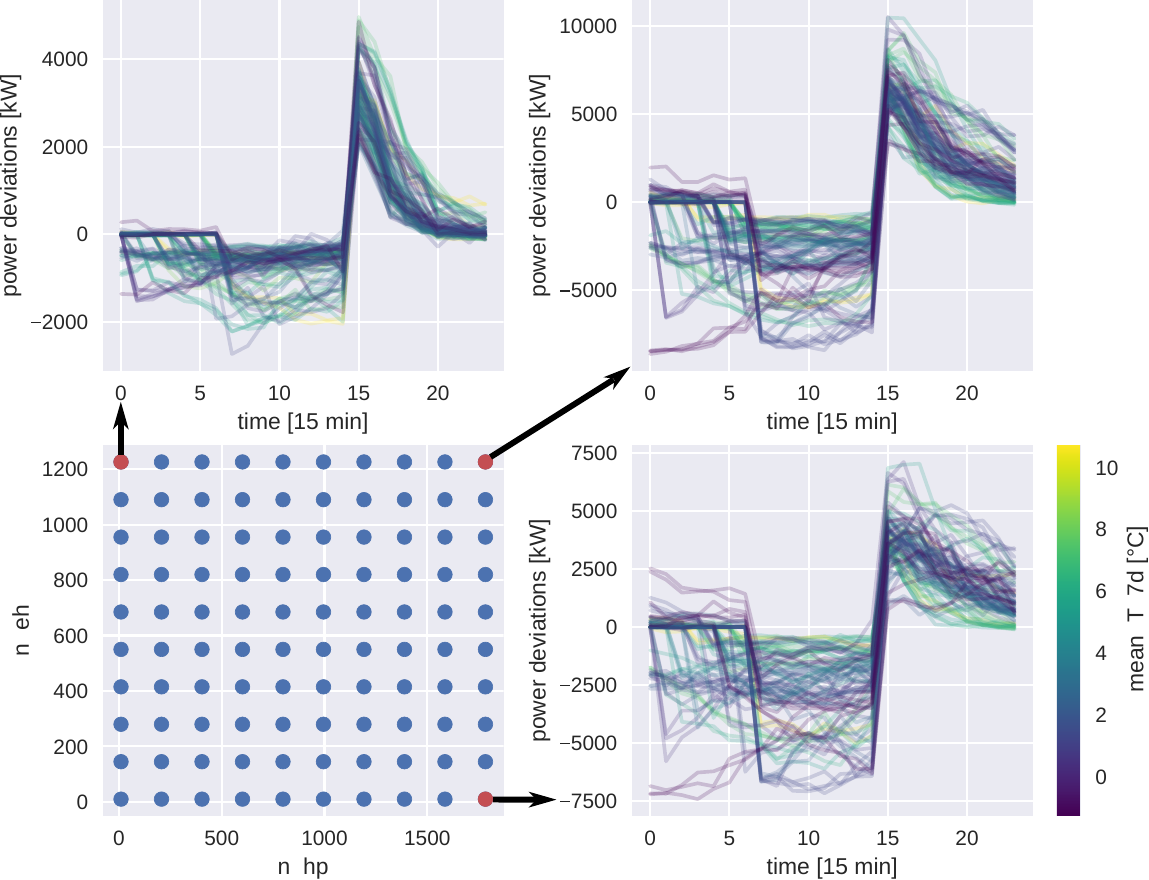}
	\caption{Example of system response in terms of deviations from the expected response (prediction where control signal features referring to feature time-steps are zeroed), dependent on the number of HPs and EHs. }
	\label{fig:rebound_examples}
\end{figure}

We used the energy imbalance aware model in combination with the grid sampling strategy to visualize rebound effects for different numbers of HPs and EHs. Figure \ref{fig:rebound_examples} shows three extreme examples of the characterization: the penetration scenario with the maximum number of EHs and zero HPs, the converse, and the scenario where both penetrations are at their maximum value. The rebound is shown in terms of energy imbalance from the test set, such that they have a force-off signal turning off at the fifteenth plotted step. It can be noticed how different observations can start to show negative energy imbalance at different time steps; this is because force-off signals can have different lengths, as shown in figure \ref{fig:scenarios}. The upper left quadrant shows the energy imbalance predicted by the metamodel in the case of the maximum number of EHs and no HPs. Comparing it with the lower right quadrant, where the sample just contains HPs, we see that the rebound effect has a quicker decay, being close to zero after only 10 steps (corresponding to 2 and a half hours). The lower right quadrant exhibits a markedly slower dissipation of the rebound effect, attributable to the different heating mechanisms and temporal constants inherent in systems heated by EHs and HPs. EHs, dedicated solely to DHW heating, have their activation guided by a hysteresis function governed by two temperature sensors installed at varying heights within the water tank. In contrast, HPs are responsible for both DHW and space heating, and their activation hinges on the temperature of the hydronic circuit, thus creating a segregation between the HPs and the building heating elements, namely the serpentine. As a result, HPs' activation is subject to a system possessing a heating capacity significantly greater than that of the standalone DHW tank: the building's heating system. Further intricacy is added to the power response profile of the heat pump due to its dual role in catering to DHW and space heating needs, with priority assigned to the former. The visual responses presented in Figure \ref{fig:scenarios} are color-differentiated according to the seven-day mean of the ambient temperature. As per the expected pattern, the EHs' responses exhibit independence from the average external temperature, while a modest influence can be detected for the HPs, where a rise in average temperatures aligns with a faster decay in response.   

\section{Using metamodels for optimal flexibility control}\label{sec:optimization}

This section presents how the metamodel can be incorporated into the optimization loop, beginning with optimizing a single flexibility group. The objective that we found most compelling from both the DSO and energy supplier perspectives is the simultaneous minimization of day-ahead costs (incurred by the energy supplier on the spot market) and peak tariff (paid by the DSO to the TSO). Notably, this scenario is particularly well-suited to Switzerland, where a distinctive situation persists with the energy supplier and the DSO remaining bundled. The peak tariff, being proportionate to the maximum monthly peak over a 15-minute interval, poses a more significant optimization challenge than day-ahead costs, as the peak tariff is paid on the monthly peak. Since it is extremely hard to produce accurate forecasts over a one-month period, we solved the peak shaving problem on a daily basis as a heuristic. This then leads us to the following optimization problem:

\begin{align}\label{eq:optimization_single}
s^* &= \argmin{s}  \mathcal{L}(\hat{y}(s)) \\
   &= \argmin{s}  \gamma\left(\sum_{h=1}^H p^s_h\hat{y}_{h}(s)\right) + p^p \max(0, \max_{h} \hat{y}_{h}(s)- y^{max}_k)  \label{eq:loss_single}
\end{align}

where $h$ refers to the step ahead, $p^s \in \mathds{R}^{T}$ is the day-ahead spot price, $p^p$ is the price for the monthly peak in $CHF/kW$, $\gamma=dt/3600$ is a coefficient taking into account the timestep duration. The second term in equation \eqref{eq:loss_single} encodes the cost of increasing the peak realized so far in the current month, $y^{max}_k$. Problem \eqref{eq:optimization_single} is not trivial to solve since it's a function of a non-parametric regressor, the metamodel. However, the parameters reported in table \ref{tab:scenarios} produce a total of 155527 control scenarios; this allows us to evaluate \eqref{eq:optimization_single} using a brute-force approach, finding the exact minimizer $s^*$. This is done through the following steps:
\begin{enumerate}
    \item Forecast the total power of the DSO: $\hat{y}^{tot} = f_{tot}(x_t, \theta_{tot})$. This forecaster was obtained by training 96 different LightGBM models, one for each step ahead.
    \item Forecast the baseline consumption of flexible devices, $\hat{y}^f(s_0) = f(x_t, s_0, \theta)$, using the metamodel with the control signal $s=s_0$ set to zero (corresponding to not controlling the devices).
    \item Forecast the response of flexible devices under a given control scenario $s$ for the next day. This is always done using the metamodel: $\hat{y}^f(s) = f(x_t, s, \theta)$. 
    \item The objective function is evaluated on $\hat{y}_t(s) = \hat{y}^{tot} -\hat{y}^f(s_0) +\hat{y}^f(s)$ for all the possible plausible control scenarios; the optimal control scenario $s^*$ minimizing the total costs is returned. 
\end{enumerate}

%\begin{figure}[!h]
%	\centering
%	\includegraphics[width=1\columnwidth]{figs/HP_forecast.png}
%	\caption{Example of forecasted signal $\hat{y}^{tot}$. Blue line: realized power. Thin lines: quantiles from 0.01 to 0.99.}
%	\label{fig:HP_forecast}
%\end{figure}

% ######## ADD Confort paragraph - dynamic grouping as function of 
\subsection{Controlling multiple groups}\label{subsec:multigroup}
\begin{figure}[!h]
	\centering
	\includegraphics[width=1\columnwidth]{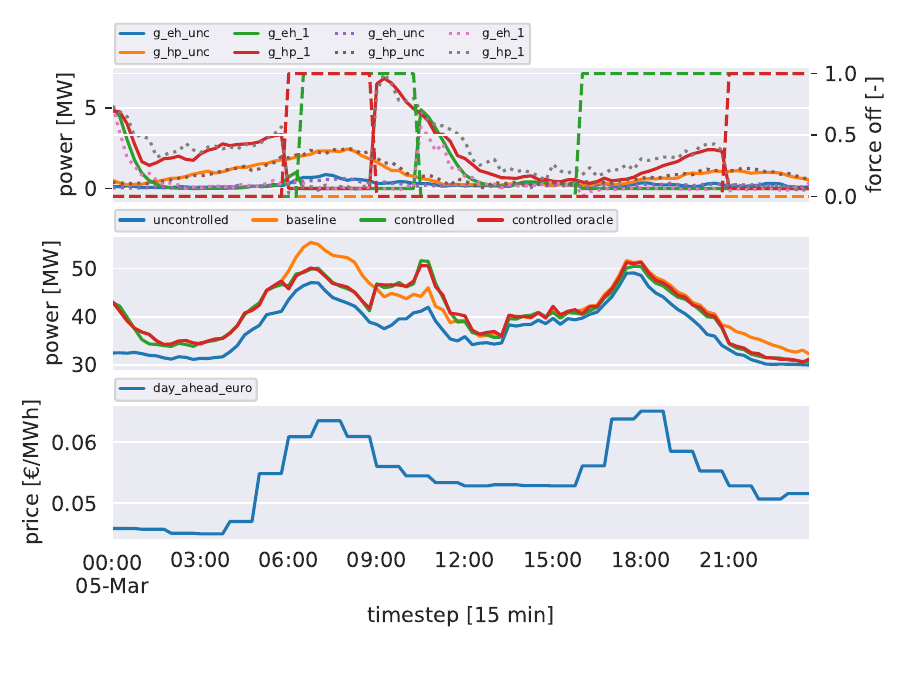}
	\caption{Example of optimized control action using the metamodel. Top: control signals (dashed), forecast group responses (dotted) and simulated, both controlled and uncontrolled, response (thick). Middle: total power from uncontrolled DSO's households (blue), total DSO's power when no control action is taken (orange), simulated and forecasted system response (green and red). Bottom: day-ahead price on the spot market.}
	\label{fig:control_example}
\end{figure}

As previously noted, forcing off a group of flexibilities results in a subsequent rebound effect when they are permitted to reactivate. A viable strategy to counter this issue is to segment the flexibilities into various groups, thereby circumventing a concurrent reactivation. Moreover, this segmentation method helps exploit their thermal inertia to the fullest extent. This is especially true in the context of heat pumps, as variations in building insulation and heating system sizing inevitably lead to differences in turn-on requirements to maintain home thermal comfort under identical weather conditions. Analogous considerations apply to hot water boilers as well. In addition, it is crucial to note that, generally, EHs can endure longer force-off periods than HPs. Thus, the stratification of flexibilities into distinct groups not only mitigates the rebound effect but also facilitates the optimal utilization of the entire appliance fleet's potential.
Problem \eqref{eq:optimization_single} can be reformulated as: 

\begin{align}\label{eq:optimization_multi}
s^* = \argmin{[s_g]_{g=1}^G} & \sum_{h=1}^H p^s_h \left(\hat{y}_h^{tot} -\sum_{g=1}^G \hat{y}_{h,g}^{f}(s_0) +\sum_{g=1}^G \hat{y}_{h, g}^{f}(s_g) \right) + \\
&p^p \max_{h} \sum_{h=1}^H \left(\hat{y}_t^{tot} -\sum_{g=1}^G \hat{y}_{h,g}^{f}(s_0) +\sum_{g=1}^G \hat{y}_{h, g}^{f}(s_g)\right)   
\end{align}
where $G$ is the total number of groups and $s_g$ is the control signal sent to the $g_{th}$ group. Problem \eqref{eq:optimization_multi} is a combinatorial problem; to reduce its complexity, we have used a sequential heuristic: the first group of devices optimizes on the uncontrolled power profile $\hat{y}_t^{tot}$. Once their optimal control for the first group is found, the second group it's optimally scheduled on $y_{t}^{tot}-\hat{y}_{t,1}^{f}(s_0) + \hat{y}_{t,1}^{f}(s)$, where the second subscript in $\hat{y}_{t,1}$ refers to the control group. An example of such sequential optimization is shown in figure \ref{fig:control_example}, where one group of EHs and one of HPs are scheduled sequentially. 

The upper panel shows the optimal control signals, along with the simulated response (dashed lines) and the response predicted by the metamodel (dotted lines). The middle panel shows the power from uncontrolled nodes in the DSO's grid (blue), the total DSO's power when no control action is taken (orange), and simulated and forecast system response (green and red).

\subsection{Ensuring comfort for the end users}\label{subsec:dynamic_grouping}
% For HPs, we chose a grouping strategy based on the energy signature of the controlled buildings. In buildings heated by a thermo-electric device, such as an HP, the energy consumption is strongly (inversely) correlated with the external temperature. The energy signature refers to a linear fit between the daily energy consumption of the building and the average daily external temperature $T_d$. Since an increasing number of households have an installed PV power plant, we also include a daily average of the global horizontal irradiance $I_d$ as a feature in the energy signature fit; high values of $I_d$ could lower the daily energy consumption if a PV plant is present, but this effect cannot be imputed to the action of temperature. Without including $I_d$ in the regression, the daily energy consumption as a function of temperature could be underestimated. The final energy signature $e(T_d, I_d)$ is a piecewise linear function of the external temperature and $I_d$. An example of an estimated energy signature is shown in figure \ref{fig:energy_signature}. 
To ensure end-user comfort while leveraging their flexibility, it is critical that appliances maintain the ability to meet energy demands for a certain period of time, despite shorter time shifts within this duration. When a building is heated with a thermo-electric device such as a heat pump (HP), its energy consumption exhibits a significant inverse correlation with the external temperature. This correlation can be effectively illustrated using an equivalent linear RC circuit to model the building's thermal dynamics.
The static behavior of this model can be represented by the energy signature, which depicts the linear relationship between the building's daily energy consumption and the mean daily external temperature, denoted as $T_d$. As more households now feature photovoltaic (PV) power plants, it becomes relevant to include the average daily global horizontal irradiance, or $I_d$, as a contributing factor in the energy signature fit. As a first approximation, we assume a linear relationship between global irradiance and PV production. Consequently, elevated $I_d$ values may correspond to lower daily energy consumption, granted a PV system is installed. However, such an effect should not be misattributed to variations in temperature. Failing to integrate $I_d$ into the regression could lead to an underestimation of the daily energy consumption when expressed as a function of temperature.
The comprehensive energy signature, denoted as $e(T_d, I_d)$, emerges as a piecewise linear function reliant on the external temperature and $I_d$. 

\begin{comment}
Figure \ref{fig:energy_signature} provides an exemplification of an estimated energy signature.

\begin{figure}[!h]
	\centering
	\includegraphics[width=1\columnwidth]{figs/energy_signature.pdf}
	\caption{Energy signature example. The average daily consumption for a household is estimated with a (piece-wise) linear function of temperature and solar irradiance. Gray dots: original observations. Crosses: fit results, colored by $I_d$}
	\label{fig:energy_signature}
\end{figure}
\end{comment}

% Finally, to retrieve the total number of activation hours $h$, we simply divide the energy signature with the nominal power:
Our ultimate objective is to ascertain the necessary operational duration for a specified HP to fulfill the building's daily energy requirements. Consequently, the total number of active hours during a day, $h$, is obtained by dividing the energy signature by the nominal power of the HP:
\begin{equation}
    h(T_d, I_d) = \frac{e(T_d, I_d)}{p_{nom}}
\end{equation}

The following steps describe our procedure to generate and control a group of HPs based on their estimated activation time:
\begin{enumerate}
\item Estimate the energy signatures of all the buildings with an installed HP $e_i(T_d, I_d)$
\item Estimate their reference activation time $h_{ref, i}$ for worst-case conditions, that is, for $T_d = 0$ and $I_d = 0$.
% \item Households are grouped based on their reference activation time: $G$ control groups are defined based on linearly spaced quantiles of $h_{ref}$.  
\item At control time, perform a day-ahead estimation of activation times for all the HPs, $h_i(\hat{T}_d, \hat{I}_d)$ using a day-ahead forecast of $T_d$ and $I_d$. Use the within-group maximum values of the needed activation time, $h_{max, g} = \max_{i \in \mathcal{G}}h_{g, i}(\hat{T}_d, \hat{I}_d)$ to filter out control scenarios having more than $h_{max, g}$ force-off steps. This process guarantees that all HPs are allowed on for a sufficient time, given the temperature and irradiance conditions.
\end{enumerate}

\section{Using metamodels for closed loop emulations}\label{sec:emulation}
For testing operational and closed-loop accuracy, we simulated one year of optimized operations, in the case in which 66\% of the available flexibilities are controlled. We used two control groups: one containing only EHs, which can be forced off for a longer period of time, and one group of HPs, controlled as explained in the previous section. 

The prediction error accuracy was already studied in section \ref{subsec:ablation}, where we tested the metamodel on a simulated test set. In that case, the force-off signals in the dataset were produced by a random policy. We further tested the performance of the metamodel when predicting the optimized force-off. We could expect a difference in prediction accuracy since, in this case, the force-off signals have a non-random pattern that could influence the average error of the forecaster. Besides this, we also assessed the accuracy of the metamodel in terms of economic results in closed-loop; that is, we retrieve the errors on the economic KPIs when the simulation is completely bypassed, and the metamodel is used for both optimizing and emulating the behavior of the controlled devices.

\subsection{Open loop operational accuracy}
\begin{figure}[!h]
	\centering
	\includegraphics[width=0.9\columnwidth]{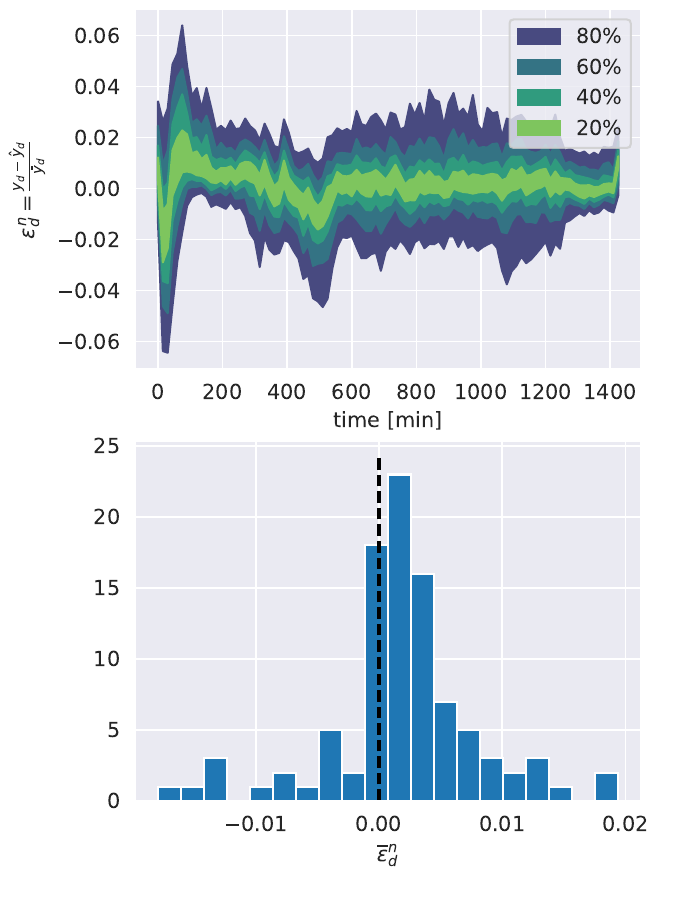}
	\caption{Performance of the metamodel in the open-loop simulations. Left: daily relative errors plotted as time series. Right: distribution of the daily means of the relative error.}
	\label{fig:oracle_openloop_perf}
\end{figure}

At first, operational accuracy was assessed in terms of predictions, comparing the aggregated controlled power profile with the sum of the individually simulated (controlled) devices. Figure \ref{fig:oracle_openloop_perf} shows the normalized daily time series of the prediction error during the actual optimization process. This is defined as:
\begin{equation}
    n\epsilon_d = \frac{y_d-\hat{y}_d}{y_d} 
\end{equation}
where $y_d, \hat{y}_d \in \mathds{R}^{96}$ are the aggregated simulated power profiles and their day ahead predictions, respectively. We see that for all the observed error paths, we just have sporadic deviations above 10\%. To have a more general understanding of the metamodel performance, in the second panel of \ref{fig:oracle_openloop_perf} we plotted the histogram of the mean daily error, defined as $\frac{1}{96} \sum_{i=1}^{96} nE_{d,i}$. This shows that the metamodel is usually under-predicting, or over-smoothing, the true response from the simulation, which is generally the expected behavior of a forecaster trained to minimize the sum of squares loss. The fact that this distribution is contained in the -2\%+2\% interval, which is much narrower than in the maximum observed discrepancies in the daily error traces, confirms that high error deviations in the day ahead predictions are just sporadic.

\subsection{Closed loop economic performances}
We cannot directly assess the closed-loop performances of the metamodel in terms of prediction errors. This is because, when simulating in a closed loop, the metamodel's predictions are fed to itself in a recurrent fashion. This could result in slightly different starting conditions for each day; furthermore, comparing the sampled paths is not our final goal. A more significant comparison is in terms of economic returns. We compared these approaches:
\begin{enumerate}
    \item Simulated: we run the optimization and fully simulate the system's response. In this setting, the metamodel is just used to obtain the optimal control signal to be applied the day ahead. The controlled devices are then simulated, subject to the optimal control signal. The costs are then computed based on the simulations.
    \item Forecast: for each day, the optimal predictions used for the optimization are used to estimate the cost. We anyways simulate the controlled devices; this process is repeated the next day. This approach gives us an understanding of how the operational prediction errors shown in figure \ref{fig:oracle_openloop_perf} impact the estimation of the costs.
    \item Emulated: the simulations are completely bypassed. The metamodel is used to optimize the control signal and generate the next-day responses for the controlled devices. 
\end{enumerate}

It should be clear that, if the third approach gives comparable results in terms of costs, we could then just use the metamodel for both the control task and its evaluation. This would significantly speed up the simulation loop: we won't have to simulate the thermodynamic behavior of thousands of households, but just evaluate the trained metamodel, which evaluation is almost instantaneous. It could seem unlikely to reach the same accuracy produced by a detailed simulation, but this can be justified by the fact that we're only interested in an aggregated power profile, whose dimensionality is just a tiny fraction of all the simulated signals needed to produce it.

\begin{figure}[!h]
	\centering
	\includegraphics[width=0.75\columnwidth]{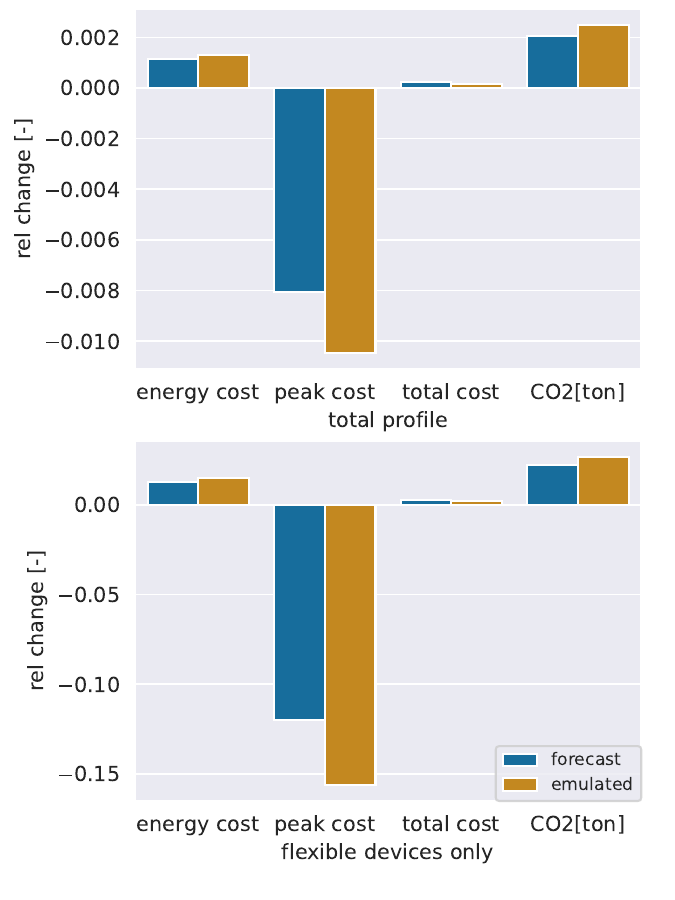}
	\caption{Deviations of different objectives from the simulated results, using the metamodel to optimize and forecast the power profiles (blue) or to completely bypass the simulation (orange). Top: relative error of objectives normalized with the total simulated costs. Bottom: relative error of objectives normalized with the additional costs faced by the DSO due to the flexible group.}
	\label{fig:closed_loop_accuracy}
\end{figure}

In figure \ref{fig:closed_loop_accuracy}, we reported the relative discrepancies from economic KPIs retrieved by the simulation, using the two aforementioned approaches. As an additional KPI, we also reported the estimated tons of produced $CO_2$. While the $CO_2$ emissions are not directly optimized for, minimizing the energy costs also positively impacts the emissions, since energy prices correlate with the $CO_2$ intensity in the energy mix. The emitted $CO_2$ tons are estimated as:
\begin{equation}
    M_{C0_2} = \sum_{t=1}^T C_t y_t
\end{equation}
where $C_t$ is the carbon intensity in the national energy mix in $\frac{g_{CO_2}}{kWh}$. 
The top panel refers to the costs that would generate considering the total power profile, $y$. In both the forecast and closed-loop cases, all costs have a deviation of less than 1\%. The total cost has a deviation of well below 1 per thousand.  In our case study, the controlled group of devices is just a small fraction of the total energy delivered by the DSO; to estimate the metamodel's performance, it's thus important to evaluate only costs generated by controlled devices $y^f$. These are shown in the bottom panel of figure \ref{fig:closed_loop_accuracy}, where we have normalized the objectives' errors with the additional costs faced by the DSO due to the flexible group:  both the energy costs and the $CO_2$ we have a relative error below the 3\%, while the peak cost has a deviation of 6\%. We have a comparable deviation for forecasts and closed-loop simulations. In all the cases, the peak costs are underestimated; this was to be expected, as the metamodel is trained with a sum of squares loss, which systematically underestimates extreme events. These discrepancies can still be considered reasonable to perform A/B testing in simulation. 

\begin{table}[h!]
\caption{\label{tab:oracle_results} First column: energy costs, peak, total costs, and $CO_2$ emissions from the controlled simulation. Second column: relative differences from the simulated costs when evaluated using the metamodel's day-ahead predictions. Third column: relative differences from the simulated costs using the metamodel to emulate the system. Data refers to the case in which 66\% of the available HPs and boilers were controlled.}
\centering
\begin{tabular}{@{}llll@{}}
\toprule
             & simulated & $\Delta_{rel}$ forecasts & $\Delta_{rel}$ closed loop\\ \midrule
Energy  & 4.18E+7 & 1.13E-3   &    1.30E-3       \\
\rowcolor[gray]{.95}Peak    & 4.46E+6 & -8.05E-3  &    -1.05E-2     \\
Total   & 4.62E+7 & 2.47E-4   &    1.68E-4     \\
\rowcolor[gray]{.95}$CO_2${[}ton{]} & 5.99E+4  & 2.06E-3   &    2.48E-3     \\ \bottomrule
\end{tabular}

\end{table}

\begin{table}[h!]
\caption{\label{tab:oracle_results_ctrl} First column: additional energy costs, peak, total costs, and $CO_2$ emissions faced by the DSO due to the flexibility group. Second and third columns as for table \ref{tab:oracle_results}}
\centering
\begin{tabular}{@{}llll@{}}
\toprule
            & simulated & $\Delta_{rel}$ forecasts & $\Delta_{rel}$ closed loop\\ \midrule
Energy  & 3.65E+6 & 1.29E-2   &    1.49E-2       \\
\rowcolor[gray]{.95}Peak    & 2.99E+5 & -1.2E-1  &    -1.56E-1     \\
Total   & 3.95E+6 & 2.88E-3   &    1.97E-3     \\
\rowcolor[gray]{.95}$CO_2${[}ton{]} & 5.58E+3  & 2.21E-2   &    2.65E-2     \\ \bottomrule
\end{tabular}
\end{table}
The left panel shows discrepancies for actual costs faced by the DSO, computed using the total power profile $y$. In this case, we have roughly a ten-fold reduction in the relative error w.r.t. the simulations. This is not a surprise, since, as anticipated, the controllable devices constitute only a fraction of the energy supplied by the DSO. Nevertheless, this is the quantity we are interested in. For completeness, the relative deviations and absolute costs for the simulated case relative to figure \ref{fig:closed_loop_accuracy} are reported in tables \ref{tab:oracle_results} and \ref{tab:oracle_results_ctrl} for the total and flexible device profiles, respectively.

\section{Conclusions and extensions}
In this work, we presented a methodology to model the flexibility potential of controllable devices located in a DSO's distribution grid and optimally steer it by broadcasting force-off signals to different clusters of flexible devices. We achieved this by training a non-parametric global forecasting model conditional to the control signals and the number of controlled devices to predict their simulated aggregated power. The numerical use case showed that the forecaster's accuracy is high enough to use it as a guide to optimally steer deferrable devices. Moreover, the high accuracy on economic KPIs suggests that the forecaster can be used to completely bypass the simulation and speed up A/B-like testing and the retrieval of different demand-side management policies over different penetration of devices.\\
We envision the following possible extensions of the presented work:
\begin{itemize}
    \item Continuous control. The presented use case relied on extensive enumeration of the possible force-off signals for the day ahead optimization. This was possible due to restrictions requested by the DSO on the shape of the control signal, which resulted in a total number of possible control signals in the order of 1e5 scenarios. Using a higher timestep for the control will require evaluating a prohibitive number of scenarios. The approach proposed in this paper can still be feasible by replacing the boosted tree with an "optimizable" regressor, that is, either a partial input-convex neural network \cite{amos_input_2017} or a conditional invertible neural network \cite{ardizzone_guided_2019}. In this case, we can use a continuous signal $s_c \in [0, 1]$ indicating the fraction of flexible devices to be forced off at a given moment in time. We can then apply gradient descent to the optimizable regressor and retrieve the optimal $s_c$.
    \item Probabilistic forecast. The presented optimization framework is based on a deterministic formulation. Formulating the problem in the stochastic framework could be advantageous when considering peak tariffs. This would require summing two sources of uncertainty: the one associated with the prediction of the total power profile $y^{
   tot}$ and the one associated with the metamodel forecasts. These can be both assessed by obtaining probability distributions after the training phase through conformal prediction and using them to generate scenarios.  
\end{itemize}

\thanks{This work was financially supported by the Swiss Federal Office of Energy (ODIS – Optimal DSO dISpatchability, SI/502074), partly by the Swiss National
Science Foundation under NCCR Automation (grant agreement 51NF40 180545), and supported by IEA Annex 82 "Energy Flexible Buildings Towards Resilient Low Carbon Energy Systems". Lorenzo Nespoli and Vasco Medici are with ISAAC, DACD, SUPSI, Mendrisio, CH (email lorenzo.nespoli@supsi.ch, vasco.medici@supsi.ch). Lorenzo Nespoli is with Hive Power SA, Manno, CH}

\bibliographystyle{ieeetr}
\bibliography{references_2}

\begin{thebibliography}{10}

\bibitem{j_cochran_et_al_flexibility_nodate}
J.~C. et~al., ``Flexibility in 2st {Century} {Power} {Systems}.''

\bibitem{babatunde_power_2020}
O.~M. Babatunde, J.~L. Munda, and Y.~Hamam, ``Power system flexibility: {A} review,'' {\em Energy Reports}, vol.~6, pp.~101--106, Feb. 2020.

\bibitem{mohandes_review_2019}
B.~Mohandes, M.~S.~E. Moursi, N.~Hatziargyriou, and S.~E. Khatib, ``A {Review} of {Power} {System} {Flexibility} {With} {High} {Penetration} of {Renewables},'' {\em IEEE Transactions on Power Systems}, vol.~34, pp.~3140--3155, July 2019.
\newblock Conference Name: IEEE Transactions on Power Systems.

\bibitem{eid_aggregation_2015}
C.~Eid, P.~Codani, Y.~Chen, Y.~Perez, and R.~Hakvoort, ``Aggregation of demand side flexibility in a smart grid: {A} review for {European} market design,'' in {\em 2015 12th {International} {Conference} on the {European} {Energy} {Market} ({EEM})}, pp.~1--5, May 2015.
\newblock ISSN: 2165-4093.

\bibitem{parvania_optimal_2013}
M.~Parvania, M.~Fotuhi-Firuzabad, and M.~Shahidehpour, ``Optimal {Demand} {Response} {Aggregation} in {Wholesale} {Electricity} {Markets},'' {\em IEEE Transactions on Smart Grid}, vol.~4, pp.~1957--1965, Dec. 2013.
\newblock Conference Name: IEEE Transactions on Smart Grid.

\bibitem{ghaemi_optimal_2019}
R.~Ghaemi, M.~Abbaszadeh, and P.~G. Bonanni, ``Optimal {Flexibility} {Control} of {Large}-{Scale} {Distributed} {Heterogeneous} {Loads} in the {Power} {Grid},'' {\em IEEE Transactions on Control of Network Systems}, vol.~6, pp.~1256--1268, Sept. 2019.
\newblock Conference Name: IEEE Transactions on Control of Network Systems.

\bibitem{noauthor_oscp_nodate}
``{OSCP} 1.0, {Protocols}, {Home} - {Open} {Charge} {Alliance}.''

\bibitem{portela_oscp-open_2015}
C.~M. Portela, P.~Klapwijk, L.~Verheijen, H.~d. Boer, and Enexis, ``{OSCP}-{An} {Open} {Protocol} {For} {Smart} {Charging} {Of} {Electric} {Vehicles},'' 2015.

\bibitem{biegel_aggregation_2014}
B.~Biegel, P.~Andersen, J.~Stoustrup, M.~B. Madsen, L.~H. Hansen, and L.~H. Rasmussen, ``Aggregation and {Control} of {Flexible} {Consumers} – {A} {Real} {Life} {Demonstration},'' {\em IFAC Proceedings Volumes}, vol.~47, no.~3, pp.~9950--9955, 2014.

\bibitem{chen_ripple-based_2016}
K.-H. Chen, ``Ripple-{Based} {Control} {Technique} {Part} {I},'' in {\em Power {Management} {Techniques} for {Integrated} {Circuit} {Design}}, pp.~170--269, IEEE, 2016.
\newblock Conference Name: Power Management Techniques for Integrated Circuit Design.

\bibitem{ponocko_forecasting_2018}
J.~Ponoćko and J.~V. Milanović, ``Forecasting {Demand} {Flexibility} of {Aggregated} {Residential} {Load} {Using} {Smart} {Meter} {Data},'' {\em IEEE Transactions on Power Systems}, vol.~33, pp.~5446--5455, Sept. 2018.
\newblock Conference Name: IEEE Transactions on Power Systems.

\bibitem{cui_evaluation_2018}
W.~Cui, Y.~Ding, H.~Hui, Z.~Lin, P.~Du, Y.~Song, and C.~Shao, ``Evaluation and {Sequential} {Dispatch} of {Operating} {Reserve} {Provided} by {Air} {Conditioners} {Considering} {Lead}–{Lag} {Rebound} {Effect},'' {\em IEEE Transactions on Power Systems}, vol.~33, pp.~6935--6950, Nov. 2018.
\newblock Conference Name: IEEE Transactions on Power Systems.

\bibitem{jensen_iea_2017}
S.~O. Jensen, A.~Marszal-Pomianowska, R.~Lollini, W.~Pasut, A.~Knotzer, P.~Engelmann, A.~Stafford, and G.~Reynders, ``{IEA} {EBC} {Annex} 67 {Energy} {Flexible} {Buildings},'' {\em Energy and Buildings}, vol.~155, pp.~25--34, Nov. 2017.

\bibitem{noauthor_iea_nodate}
``{IEA} {EBC} {\textbar}{\textbar} {Annex} 82 {\textbar}{\textbar} {Energy} {Flexible} {Buildings} {Towards} {Resilient} {Low} {Carbon} {Energy} {Systems} {\textbar}{\textbar} {IEA} {EBC} {\textbar}{\textbar} {Annex} 82.''

\bibitem{six2011exploring}
D.~Six, J.~Desmedt, D.~Vahnoudt, and J.~Bael, ``Exploring the flexibility potential of residential heat pumps combined with thermal energy storage for smart grids,'' in {\em 21th international conference on electricity distribution, paper}, vol.~442, 2011.

\bibitem{chen_measures_2018}
Y.~Chen, P.~Xu, J.~Gu, F.~Schmidt, and W.~Li, ``Measures to improve energy demand flexibility in buildings for demand response ({DR}): {A} review,'' {\em Energy and Buildings}, vol.~177, pp.~125--139, Oct. 2018.

\bibitem{balint_determinants_2019}
A.~Balint and H.~Kazmi, ``Determinants of energy flexibility in residential hot water systems,'' {\em Energy and Buildings}, vol.~188-189, pp.~286--296, Apr. 2019.

\bibitem{junker_characterizing_2018}
R.~G. Junker, A.~G. Azar, R.~A. Lopes, K.~B. Lindberg, G.~Reynders, R.~Relan, and H.~Madsen, ``Characterizing the energy flexibility of buildings and districts,'' {\em Applied Energy}, vol.~225, pp.~175--182, Sept. 2018.

\bibitem{nuytten_flexibility_2013}
T.~Nuytten, B.~Claessens, K.~Paredis, J.~Van~Bael, and D.~Six, ``Flexibility of a combined heat and power system with thermal energy storage for district heating,'' {\em Applied Energy}, vol.~104, pp.~583--591, Apr. 2013.

\bibitem{petersen_taxonomy_2013}
M.~K. Petersen, K.~Edlund, L.~H. Hansen, J.~Bendtsen, and J.~Stoustrup, ``A taxonomy for modeling flexibility and a computationally efficient algorithm for dispatch in {Smart} {Grids},'' in {\em 2013 {American} {Control} {Conference}}, pp.~1150--1156, June 2013.
\newblock ISSN: 2378-5861.

\bibitem{de_coninck_quantification_2016}
R.~De~Coninck and L.~Helsen, ``Quantification of flexibility in buildings by cost curves – {Methodology} and application,'' {\em Applied Energy}, vol.~162, pp.~653--665, Jan. 2016.

\bibitem{oldewurtel_towards_2013}
F.~Oldewurtel, D.~Sturzenegger, G.~Andersson, M.~Morari, and R.~S. Smith, ``Towards a standardized building assessment for demand response,'' in {\em 52nd {IEEE} {Conference} on {Decision} and {Control}}, pp.~7083--7088, Dec. 2013.
\newblock ISSN: 0191-2216.

\bibitem{reynders_generic_2017}
G.~Reynders, J.~Diriken, and D.~Saelens, ``Generic characterization method for energy flexibility: {Applied} to structural thermal storage in residential buildings,'' {\em Applied Energy}, vol.~198, pp.~192--202, July 2017.

\bibitem{fischer_model-based_2017}
D.~Fischer, T.~Wolf, J.~Wapler, R.~Hollinger, and H.~Madani, ``Model-based flexibility assessment of a residential heat pump pool,'' {\em Energy}, vol.~118, pp.~853--864, Jan. 2017.

\bibitem{muller_large-scale_2019}
F.~Müller and B.~Jansen, ``Large-scale demonstration of precise demand response provided by residential heat pumps,'' {\em Applied Energy}, vol.~239, pp.~836--845, Apr. 2019.

\bibitem{valles_probabilistic_2018}
M.~Vallés, A.~Bello, J.~Reneses, and P.~Frías, ``Probabilistic characterization of electricity consumer responsiveness to economic incentives,'' {\em Applied Energy}, vol.~216, pp.~296--310, Apr. 2018.

\bibitem{corradi_controlling_2013}
O.~Corradi, H.~Ochsenfeld, H.~Madsen, and P.~Pinson, ``Controlling {Electricity} {Consumption} by {Forecasting} its {Response} to {Varying} {Prices},'' {\em IEEE Transactions on Power Systems}, vol.~28, pp.~421--429, Feb. 2013.
\newblock Conference Name: IEEE Transactions on Power Systems.

\bibitem{junker_stochastic_2020}
R.~G. Junker, C.~S. Kallesøe, J.~P. Real, B.~Howard, R.~A. Lopes, and H.~Madsen, ``Stochastic nonlinear modelling and application of price-based energy flexibility,'' {\em Applied Energy}, vol.~275, p.~115096, Oct. 2020.

\bibitem{yin_long-term_2022}
L.~Yin and Y.~Qiu, ``Long-term price guidance mechanism of flexible energy service providers based on stochastic differential methods,'' {\em Energy}, vol.~238, p.~121818, Jan. 2022.

\bibitem{noauthor_welcome_nodate}
``Welcome to {LightGBM}’s documentation! — {LightGBM} 3.3.1.99 documentation.''

\bibitem{noauthor_optuna_nodate}
``Optuna - {A} hyperparameter optimization framework.''

\bibitem{ozaki_multiobjective_2020}
Y.~Ozaki, Y.~Tanigaki, S.~Watanabe, and M.~Onishi, ``Multiobjective tree-structured parzen estimator for computationally expensive optimization problems,'' in {\em Proceedings of the 2020 {Genetic} and {Evolutionary} {Computation} {Conference}}, {GECCO} '20, (New York, NY, USA), pp.~533--541, Association for Computing Machinery, June 2020.

\bibitem{amos_input_2017}
B.~Amos, L.~Xu, and J.~Z. Kolter, ``Input {Convex} {Neural} {Networks},'' {\em Proceedings of the 34 th International Conference on Machine Learning}, p.~10, 2017.

\bibitem{ardizzone_guided_2019}
L.~Ardizzone, C.~Lüth, J.~Kruse, C.~Rother, and U.~Köthe, ``Guided {Image} {Generation} with {Conditional} {Invertible} {Neural} {Networks},'' July 2019.
\newblock arXiv:1907.02392 [cs].

\bibitem{bergman_fundamentals_2011}
T.~L. Bergman, F.~P. Incropera, D.~P. DeWitt, and A.~S. Lavine, {\em Fundamentals of {Heat} and {Mass} {Transfer}}.
\newblock John Wiley \& Sons, Apr. 2011.
\newblock Google-Books-ID: vvyIoXEywMoC.

\bibitem{cholewa_heat_2013}
T.~Cholewa, M.~Rosiński, Z.~Spik, M.~R. Dudzińska, and A.~Siuta-Olcha, ``On the heat transfer coefficients between heated/cooled radiant floor and room,'' {\em Energy and Buildings}, vol.~66, pp.~599--606, Nov. 2013.

\bibitem{nespoli_distributed_2017}
L.~Nespoli, A.~Giusti, N.~Vermes, M.~Derboni, A.~E. Rizzoli, L.~M. Gambardella, and V.~Medici, ``Distributed demand side management using electric boilers,'' {\em Computer Science - Research and Development}, vol.~32, pp.~35--47, Mar. 2017.

\bibitem{of_topography_mapgeoadmin_nodate}
F.~O. of~Topography, ``Mapgeoadmin.''

\bibitem{federal_statistical_office_ch_statistica_2016}
C.~Federal Statistical~Office, ``Statistica degli edifici e delle abitazioni (dal 2009) {\textbar} {Fact} sheet,'' Oct. 2016.

\bibitem{pampuri_evaluation_2017}
L.~Pampuri, N.~Cereghetti, P.~G. Bianchi, and P.~Caputo, ``Evaluation of the space heating need in residential buildings at territorial scale: {The} case of {Canton} {Ticino} ({CH}),'' {\em Energy and Buildings}, vol.~148, pp.~218--227, Aug. 2017.

\bibitem{bruttisellen-zurich_sia-shop_nodate}
Brüttisellen-Zurich, ``{SIA}-{Shop} {Produkt} - {SIA} 380 / 2015 {I} - {Basi} per il calcolo energetico di edifici.''

\bibitem{streicher_analysis_2019}
K.~N. Streicher, P.~Padey, D.~Parra, M.~C. Bürer, S.~Schneider, and M.~K. Patel, ``Analysis of space heating demand in the {Swiss} residential building stock: {Element}-based bottom-up model of archetype buildings,'' {\em Energy and Buildings}, vol.~184, pp.~300--322, Feb. 2019.

\end{thebibliography}

\FloatBarrier
\appendix
\section{Detailed simulation's model}\label{annex_1}
\subsection{HP control logic}\label{ann:control_heating}
The heating system is modeled using the STASCH 6 standard.
The heat pump control logic is based on two temperature sensors placed at different heights of the water tank, while the circulation pump connecting the tank with the building's heating element is controlled by a hysteresis on the temperature measured by a sensor placed inside the house. \\
We describe the control logic in a sequential way, following the heating components of the system. The first decision is taken by the building central controller, which decides its working mode, that is, if the building needs to be cooled or heated, based on a moving average of the historical data of the external temperature:
\begin{equation}
\begin{cases}
wm_t&= -1 \qquad \textbf{if} \quad T_{ma,t}>T_{max,ma}\\
wm_t&= \ \ 1 \qquad \textbf{if} \quad T_{ma,t}<T_{min,ma}\\ 
wm_t&= \ \ 0 \qquad otherwise
\end{cases}
\end{equation}
where the working mode $wm_t$ is negative when the building requires to be cooled, positive when heating is required, and 0 when no actions are needed.$T_{max,ma}$ and $T_{min,ma}$ represent the maximum and minimum values of the external temperature's moving average, which is based on the past 7 days.
The actual activation of the heating element is controlled by the hysteresis on the internal temperature of the building, $T_z$. If the working mode is positive, this is given by:
\begin{equation}\label{eq:heating_hy}
\begin{cases}
s_{hy,t}&= 1 \qquad \textbf{if} \quad \left(\ T_z<T_{min,hy}-\Delta T/2\right) \\
& \qquad \quad \ \textbf{or} \quad  \left(T_z < T_{min,hy}+\Delta T/2 \quad \textbf{and} \quad s_{hy,t-1}\right)\\
s_{hy}&=  0 \qquad otherwise\\ 
\end{cases}
\end{equation}
where $s_{hy,t}$ is the state of the hysteresis at time $t$, 1 meaning that the circulation pump of the heating element must be activated, and $\Delta T$ was chosen to be equal to 1$^\circ C$. For completeness, we report also the control logic when the building is in cooling mode:
\begin{equation}
\begin{cases}
s_{hy,t}&= 1 \qquad \textbf{if} \quad \left(\ T_z>T_{max,hy}+\Delta T/2\right) \\
& \qquad \quad \ \textbf{or} \quad  \left(T_z > T_{max,hy}-\Delta T/2 \quad \textbf{and} \quad s_{hy,t-1}\right)\\
s_{hy}&=  0 \qquad otherwise\\ 
\end{cases}
\end{equation}
The incoming water temperature in the heating element is then modulated linearly through a 3-way valve between a maximum and minimum value, based on the external temperature, both in the heating and cooling modes.
When operative, the heating element requests hot or cold water to the water tank, which control logic is based on two temperature sensors located in two different layers. When the building is in heating mode, the control logic is a simple hysteresis based on the temperature of the sensor in the uppermost layer, which is identical to the one in \eqref{eq:heating_hy}. When in cooling mode, the control logic is the following:
\begin{equation}
\begin{cases}
s_{hy,t}&= -1 \qquad \textbf{if} \quad \left(\ T_{up}>T_{max}^c+\Delta T/2\right) \\
& \qquad \qquad \  \textbf{or} \quad  T_{low} > T_{max}^c+\Delta T/2 \\
s_{hy,t}&=  0 \qquad \quad \textbf{if} \quad \left(\ T_{low}<T_{min}^c \right) \ \textbf{or} \left(T_{up}<T_{max}^c-\Delta T /2\right)\\ 
s_{hy,t}&= s_{hy,t-1} \quad otherwise
\end{cases}
\end{equation}
where $T_{up}$ and $T_{low}$ are the temperature measured by the upper and lower sensors, respectively, and $T_{min}^c$ and $T_{max}^c$ are the minimum and maximum desired temperatures of the water in the tank while in cooling mode. \\
The value of $s_{hy,t}$ is then communicated to the HP. In the case in which the HP is also used for the domestic hot water (DHW), the DHW tank is always served with priority by the HP.  

\subsection{Heat distribution system}\label{ann:heat_distribution}
Floor heating was modeled starting from the first principles. Considering a fixed and uniform temperature for the ground and the building internal temperature at each time-step and stationary conditions, we can retrieve the analytical expression of the temperature profile along the pipe, through the energy balance on an infinitesimal element of the pipe. This can be expressed as:
\begin{equation}
\frac{\partial c T_x}{\partial t} = \Phi_x -\Phi_{x+\partial x} + \dot{q}_{up} + \dot{q}_{down}
\end{equation}    
where $c$ is the heat capacity in $J/K$,  $x$ is the distance from the pipe entrance, $T_x$ is the temperature of the water inside the pipe at $x$, $\Phi$ are enthalpy flows at the entrance and exit of the considered infinitesimal volume, $\dot{q}_{up}$ and $\dot{q}_{down}$ are the heating powers from the building and from the ground.
Expressing the latter through equivalent resistance taking into account convective and conductive effects, the balance in steady state can be rewritten as:
\begin{equation}\label{eq:floor_heating}
\frac{\dot{m}c_p}{\rho^*} \frac{\partial T_x}{\partial x} = \frac{R_{down}T_z +R_{up}T_g}{R_{down}+R_{up}} -T_x = T^a-T_x
\end{equation}  
where $T^a$ is the asymptotic temperature and where:
\begin{align}
R_{down}&= \frac{1}{h_{in} w} + \frac{1}{h_{u,eq} w} + R_u\\
R_{up}&=\frac{1}{h_{in} w} + R_g\\
\rho^* &= \frac{R_{up}+R_{down}}{R_{up}R_{down}}
\end{align}
where $w$ is the diameter of the tube, $h_{in}$ is the internal coefficient of heat transfer, which can be retrieved using available empirical relation for fully developed flow with fixed temperature at the boundary conditions \cite{bergman_fundamentals_2011}, $h_{u,eq}$ is the heat transfer coefficient between the floor and the building air including both the effect for natural convection and radiation. The values of $h_{u,eq}$ can be found in the literature \cite{cholewa_heat_2013}. The value of the thermal resistances $R_u$ and $R_g$, towards the floor and the ground, can be found in the literature as well. We can reformulate \eqref{eq:floor_heating}, making it adimensional through a change of variable:

\begin{equation}
\frac{\partial\Theta}{\partial\mathcal{X}} = -\Theta
\end{equation}

from which solution we can retrieve the temperature profile of the water inside the pipe:
\begin{equation}\label{eq:T_profile}
T_x = T^a + (T_0 -T^a)e^{\frac{-x\rho^*}{\dot{m}c_p}}
\end{equation}
where $T_0$ is the temperature of the water at the pipe inlet. We can use \eqref{eq:T_profile} to retrieve the heating power flowing into the building, integrating $\dot{q}_{up}(x)$ along the pipe.
\begin{equation}
\dot{Q}_{up} = \int_{0}^{L}\dot{q}_{up}(x)\mathrm{d}x=\int_{0}^{L}\frac{T(x)-T_z}{R_{up}}\mathrm{d}x
\end{equation}
where $L$ is the length of the serpentine. Integrating, we obtain
\begin{equation}\label{eq:qup}
\dot{Q}_{up} = \frac{(T^a-T_z)L-(T_L-T_0)\frac{\dot{m}c_p}{\rho^*}}{R_{up}}
\end{equation}
where $T_L$ is the temperature of the water at the outlet of the serpentine. Note that the equation \eqref{eq:qup} tends to $(T_L-T_0)\dot{m}c_p$ when $R_{down}$ increase and $R_{up}$ is kept fixed.\\
The nominal mass flow of the heating system and the length of the serpentine are found as the solution of the following optimization problem:
\begin{equation}
\argmin{L,\dot{m}} \left(\dot{Q}_{up}(L)-\dot{Q}_{nom}\right)^2 + 10^{-3}\left(\dot{m}-\dot{m}_{nom}\right)^2
\end{equation}
where $\dot{m}_{nom}$ is a reference mass flow, equal to $0.1 \left[kg/s\right]$ and $\dot{Q}_{nom}$ is the power required to keep the building internal temperature constant under reference conditions (we used an external temperature of -4$^\circ C$ and a desired internal temperature of 20 $^\circ C$):
\begin{equation}
\dot{Q}_{nom} = \frac{\Delta T_{ref}}{R}
\end{equation}
where $R$ is the resistance of an equivalent RC circuit describing the heating dynamics of the building.

\subsection{Water tank model}\label{ann:water_tank}
The dynamic equation describing the evolution of the temperature of the tank's layers is the following:
\begin{align}
C \frac{\partial T_i}{\partial t} &= \dot{Q}_{buo,i}^u + \dot{Q}_{buo,i}^d + \dot{Q}_{h,i}+ \dot{Q}_{loss,i} \\ \nonumber
& + \dot{Q}_{cond,i}^u + \dot{Q}_{cond,i}^d + c_p \dot{m} (T_{i-1} - T_i)
\end{align}
where $T_i$ is the temperature of the $i_{th}$ layer, $Q_{buo}^u$,$Q_{buo}^d$,$Q_{cond}^u$,$Q_{cond}^u$ are the thermal powers due to buoyancy and conduction, from the lower and upper layer, respectively. The last term represents the enthalpy flow due to mass exchange, while $C$ is the thermal capacity of the layer, in $[J/K]$ and $Q_{h,i}$ is the thermal power due to an electric resistance (for the boiler) or an heat exchange (for the heating system buffer).
The expression for the above thermal power are the following:
\begin{align}
\dot{Q}_{buo,i}^u &= k \ \mathrm{max}(T_{i+1}-T_i, 0)N, \quad 0 \quad for \quad i = N\\
\dot{Q}_{buo,i}^d &=  k \ \mathrm{max}(T_{i-1}-T_i, 0)N, \quad 0 \quad for \quad i = 1   \\
\dot{Q}_{cond,i}^u &= u_{amb}(T_{i+1}-T_i), \quad 0 \quad for \quad i = N\\
\dot{Q}_{cond,i}^d &= u_{amb}(T_{i-1}-T_i), \quad 0 \quad for \quad i = 1\\
\dot{Q}_{loss,i} &= u_{amb}(T_{ext}-T_{i})\\
\dot{Q}_{h,i} &= \dot{Q}_{tot}/n_{h} \quad if \quad i \in \mathcal{I}\\
\end{align}
where $N$ is the number of layers, $u_{amb}$ is the equivalent thermal loss coefficient with the ambient and $\mathcal{I}$ is the set of the $n_h$ layers heated by the heat exchange (or electric resistance). The buoyancy model is the one proposed in the
IDEAS library \cite{de_coninck_quantification_2016}.
A detailed description of the parameters for the boiler model can be found in \cite{nespoli_distributed_2017}.

\section{Metadata sources}\label{annex_2}
\begin{figure}[h]
	\centering
	\includegraphics[width=1\columnwidth]{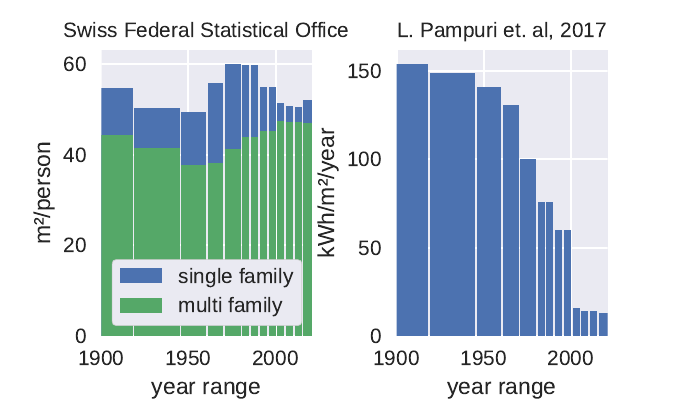}
	\caption{Representative values of m$^2$/person (Switzerland) and kWh/m$^2$/year (Switzerland, canton Ticino) for buildings, conditional to the class of construction year.}\label{fig:stats_squared_meters_consumption}
\end{figure}
To faithfully simulate the system, we need to estimate the presence of an HP or EH, the number of dwellers (influencing DHW consumption), and the equivalent thermal resistance $R\ [\qty{}{\kelvin\per\watt}]$ and capacity $C \ [\qty{}{\kilo\watt h\per\kelvin}]$ of buildings. We retrieved information on which building is equipped with an HP or an EH in a given region using data from \cite{of_topography_mapgeoadmin_nodate}, \cite{federal_statistical_office_ch_statistica_2016}. We then combine this information with the following, summarized in figure \ref{fig:stats_squared_meters_consumption}:
\begin{itemize}
    \item the average number of m$^2$ per person for buildings of a given construction age, from \cite{federal_statistical_office_ch_statistica_2016}, which allows us to have an estimate of the number of dwellers. This information is then used to retrieve a water consumption profile and to size the heating source and buffer volume for the DHW.
    \item the total annual consumption per square meter and construction age of buildings in the region, from \cite{pampuri_evaluation_2017}, and the heating reference surface (HRS) from \cite{of_topography_mapgeoadmin_nodate}, which are then used to estimate the equivalent building's thermal resistance $R$.
\end{itemize}

A summary of the final set of parameters, the conditioning factors, and the sources used to retrieve them is reported in table \ref{tab:sim_parameters}.

\begin{table}[]
\caption{Upper and lower bounds for the uniform distribution for the sizing of the EH\label{tab:eh_params}}
	\centering
	\begin{tabular}{@{}cccc@{}}
		\toprule
		\multicolumn{2}{c}{power {[}kW/person{]}}                           & \multicolumn{2}{c}{volume {[}m$^3$/person{]}}                          \\ \midrule
		\multicolumn{1}{c|}{min} & \multicolumn{1}{c|}{max} & \multicolumn{1}{c|}{min} & \multicolumn{1}{c}{max} \\ \midrule
		1                                 & 2                                & 0.08                             & 0.12                             \\ \bottomrule
	\end{tabular}
\end{table}
\begin{table}[h]
\caption{Simulation parameters and their sources\label{tab:sim_parameters}}
    \begin{tabular}{ccc}
        \toprule
        parameter            & conditional on & sources \\
        \hline
        $R\ [\qty{}{\kelvin\per\watt}]$       & \makecell{construction period, location,\\ class of building}    & \cite{of_topography_mapgeoadmin_nodate, pampuri_evaluation_2017} \\
        \rowcolor[gray]{.95}  $C \ [\qty{}{\kilo\watt h\per\kelvin}]$ & - & \cite{bruttisellen-zurich_sia-shop_nodate} \\
        Prob(HP - EH) & \makecell{construction period, location,\\ class of building} &  \cite{of_topography_mapgeoadmin_nodate, streicher_analysis_2019} \\
        \rowcolor[gray]{.95}occupants & \makecell{construction period, location,\\ class of building} &  \cite{of_topography_mapgeoadmin_nodate, federal_statistical_office_ch_statistica_2016}\\
        \bottomrule
    \end{tabular}
\end{table}








\end{document}